\documentclass[
superscriptaddress,
nofootinbib,
 amsmath,amssymb,
 aps,
floatfix,
prd,
showkeys,
10pt,
twocolumn,
]{revtex4-2}

\usepackage{xcolor}

\renewcommand{\thefootnote}{\fnsymbol{footnote}}
\usepackage{bbm}
\usepackage{ulem}
\usepackage{tabularx}
\usepackage{graphicx}
\usepackage{amsmath}
\usepackage{hyperref}
\hypersetup{
    colorlinks,
    linkcolor={red!50!black},
    citecolor={blue!50!black},
    urlcolor={blue!80!black}
}

\graphicspath{{./}{figures/}}

\begin{document}

\title{Investigating the Impact of Higher-Order Phase Transitions in Binary Neutron-Star Mergers}

\author{P. Hammond}
\affiliation{Institute for Gravitation and the Cosmos, The Pennsylvania State University, University Park, PA 16802, USA}
\affiliation{Department of Physics, The Pennsylvania State University, University Park, PA 16802, USA}
\affiliation{Department of Physics and Astronomy, University of New Hampshire, Durham, NH 03824, USA}

\author{A. Clevinger}
\affiliation{Center for Nuclear Research, Department of Physics, Kent State University, Kent, OH 44242 USA}

\author{M. Albino}
\affiliation{CFisUC, Department of Physics, University of Coimbra, 3004-516 Coimbra, Portugal}

\author{V. Dexheimer}
\email{vdexheim@kent.edu}
\affiliation{Center for Nuclear Research, Department of Physics, Kent State University, Kent, OH 44242 USA}

\author{S. Bernuzzi}
\affiliation{Theoretisch-Physikalisches Institut, Friedrich-Schiller-Universit\"{a}t Jena, 07743, Jena, Germany}

\author{C. Brown}
\affiliation{Center for Nuclear Research, Department of Physics, Kent State University, Kent, OH 44242 USA}

\author{W. Cook}
\affiliation{Theoretisch-Physikalisches Institut, Friedrich-Schiller-Universit\"{a}t Jena, 07743, Jena, Germany}

\author{B. Daszuta}
\affiliation{Theoretisch-Physikalisches Institut, Friedrich-Schiller-Universit\"{a}t Jena, 07743, Jena, Germany}

\author{J. Fields}
\affiliation{Institute for Gravitation and the Cosmos, The Pennsylvania State University, University Park, PA 16802, USA}
\affiliation{Department of Physics, The Pennsylvania State University, University Park, PA 16802, USA}

\author{E. Grundy}
\affiliation{Center for Nuclear Research, Department of Physics, Kent State University, Kent, OH 44242 USA}

\author{C. Providência}
\affiliation{CFisUC, Department of Physics, University of Coimbra, 3004-516 Coimbra, Portugal}

\author{D. Radice}
\affiliation{Institute for Gravitation and the Cosmos, The Pennsylvania State University, University Park, PA 16802, USA}
\affiliation{Department of Physics, The Pennsylvania State University, University Park, PA 16802, USA}
\affiliation{Department of Astronomy \& Astrophysics, The Pennsylvania State University, University Park, PA 16802, USA}

\author{A. Steiner}
\affiliation{Department of Physics and Astronomy, University of Tennessee, Knoxville, TN 37996, USA}
\affiliation{Physics Division, Oak Ridge National Laboratory, Oak Ridge, TN 37831, USA}

\def\ls{\textcolor{purple}}

\begin{abstract}
In this paper we investigate quark deconfinement in neutrons stars and their mergers, focusing on the effects of higher orders for the phase transition between hadronic and quark matter. The different  descriptions we use to describe matter microscopically contain varying particle degrees of freedom, including nucleons, hyperons, Delta baryons, and light and strange quarks. We use tabulated equations of state from the CompOSE database in which the quark deconfinement phase transition is described as being first-order, and then smooth it out by introducing a percolation, replacing the single first-order phase transition with two transitions of second or third order. We then perform binary neutron-star merger simulations using these new equations of st ate, focusing on groups of binaries with the same single-star mass, radius, and tidal deformability, but different equations of state. We go on to discuss differences in their evolution, and the ramifications for interpreting future gravitational wave observations and the potential to learn about dense matter. 
\end{abstract}   


\maketitle

\section{Introduction}

Since the Laser Interferometer Gravitational-Wave Observatory (LIGO) first started detecting gravitational waves in 2015 \cite{LIGOScientific:2016aoc}, a whole new way to observe neutron stars has emerged. Several binary neutron-star (BNS) merger events have already been detected \cite{LIGOScientific:2017vwq,LIGOScientific:2017ync,LIGOScientific:2017zic,LIGOScientific:2018hze,LIGOScientific:2020aai} and we expect to start observing more of these events as the sensitivity of LIGO increases in subsequent observation periods and as third generation gravitational wave detectors (\cite{Punturo:2010zz, Hild:2010id, ET:2019dnz,Reitze:2019iox,KAGRA:2013rdx,Evans:2021gyd,Kiendrebeogo:2023hzf,Capote:2024rmo}~\footnote{See also The Einstein Telescope: \url{http://www.et-gw.eu/}, Cosmic Explorer: \url{https://cosmicexplorer.org/}.}) come online. Gravitational waves emitted by neutron-star mergers are relatively weak in comparison to binary black hole (BBH) mergers, so closer sources are required for detection \citep{KAGRA:2013rdx}, which, combined with population estimates, results in a lower expected observation rate for BNSs as compared to BBHs \cite{LIGOScientific:2018mvr,Kiendrebeogo:2023hzf}. Nonetheless, gravitational-wave signals from neutron-star mergers are important \cite{Cutler:1992tc} because they can provide us with unique insight into the interior and, consequently, the composition of the core of neutron stars (see, e.g., \cite{Lattimer:2000nx,Lattimer:2004pg,Hinderer:2009ca, Damour:2012yf,Sekiguchi:2011mc,Hotokezaka:2011dh,Bauswein:2013jpa,Lattimer:2015nhk,Radice:2016rys,Margalit:2017dij,Bauswein:2017vtn,Radice:2017lry,LIGOScientific:2017fdd,Baym:2017whm,Most:2018eaw,Bauswein:2018bma,Coughlin:2018fis,De:2018uhw,LIGOScientific:2018hze,LIGOScientific:2018cki,Radice:2018ozg,Most:2019onn,Dietrich:2020efo,Lattimer:2021emm,Breschi:2021tbm,Breschi:2021xrx,Kashyap:2021wzs,Perego:2021mkd,Prakash:2021wpz,Fujimoto:2022xhv,Chatziioannou:2024tjq,Koehn:2024set,Annala:2019puf,Annala:2023cwx,Prakash:2021wpz,Prakash:2023afe,Espino:2023llj,Haque:2022dsc,Blacker:2020nlq,Most:2022wgo,Huang:2022mqp,Kedia:2022nns,Liebling:2020dhf,Weih:2019xvw,Nandi:2017rhy,Zhao:2017nlw,Blacker:2020nlq,Haque:2022dsc,Banik:2017zia,Kiuchi:2019kzt,Nunna:2020jzt,Bombaci:2020vgw,Matur:2024nwi,Gieg:2024jxs,Callister:2024cdx,Kochankovski:2024lqn,Ng:2024zve,Kochankovski:2025lqc,Burgio:2018yix,Alvarez-Castillo:2016oln,Bauswein:2020ggy,Demircik:2020jkc,Blacker:2024tet,Hensh:2024onv,Chatterjee:2025zrh,Counsell:2025hcv}).
Inside neutron stars, we expect baryon densities several times saturation density, $n_0$, which cannot be produced in terrestrial experiments (without large contribution from temperature). In this unique regime, it is likely for quark deconfinement to occur (see, e.g., \cite{Most:2018eaw, Most:2019onn, Bauswein:2018bma,Annala:2019puf, Prakash:2021wpz,Annala:2023cwx,Burgio:2018yix,Alvarez-Castillo:2018pve,Fujimoto:2022xhv,Chatterjee:2025zrh}) and more so when neutron stars merge.

The deconfinement phase transition could be first order, which would imply a discontinuity in (baryon number) density, $n_\mathrm{B}$, or it could be of higher order, which would imply discontinuities in higher order derivatives of the pressure $P$, $\mathrm{d}^n P/\mathrm{d} \mu_\mathrm{B}^n$, where $n$ is the order of the phase transition and $\mu_\mathrm{B}$ the baryon chemical potential, such that $n_\mathrm{B}=\mathrm{d} P/ \mathrm{d}\mu_\mathrm{B}$. In the case that all of these derivatives are continuous, that would imply a crossover. Evidence of first-order phase transitions could possibly be seen in gravitational-wave signals from neutron-star mergers \cite{Most:2018eaw,Bauswein:2018bma,Liebling:2020dhf,Prakash:2021wpz,Prakash:2023afe,Espino:2023llj,Chatziioannou:2019yko,Demircik:2020jkc,Blacker:2024tet,Hensh:2024onv,Counsell:2025hcv}. There have also been studies on mergers of neutron stars described by equations of state (EoSs) with mixed phases \cite{Blacker:2020nlq,Prakash:2021wpz,Haque:2022dsc,Prakash:2023afe}, as well as EoSs with higher-order phase transitions \cite{Most:2022wgo,Huang:2022mqp,Kedia:2022nns,Weih:2019xvw, Nandi:2017rhy, Blacker:2020nlq,Haque:2022dsc,Hensh:2024onv}. Note that a first-order phase transition with a mixture of phases differs from a second-order phase transition. Although both cases present discontinuities in $\mathrm{d}^2 P / \mathrm{d} \mu_\mathrm{B}^2$, the former presents two discontinuities (around the mixed phase, with continuously varied charge chemical potential, $\mu_\mathrm{Q}$), while the later presents one (at the transition).

Within first-order phase transitions, the speed of sound, $c_s = \sqrt{\mathrm{d} P / \mathrm{d} \varepsilon}$ drops to zero in the range of baryon number and energy densities, $n_\mathrm{B}$ and $\varepsilon$, of the phase transition. In this work we alter the speed of sound around the phase transition region by means of a percolation, as suggested in Ref.~\cite{Baym:1979etb,Masuda:2012kf,Kojo:2014rca}. This procedure is based on the hypothesis that the quarks become deconfined but still remain localized for a certain regime of densities, giving rise to a new phase of matter called the \textit{quarkyonic} phase \cite{McLerran:2007qj}. We note that percolation produces EoSs that can appear similar to the mixed-phase EoSs previously mentioned---there are two regimes bracketing some intermediate regime---but are fundamentally different, in that percolation introduces a new third phase, rather than some representation of a combination/coexistence of the other two. Additionally, we have significantly more control over the position and order of the transitions to/from the percolation phase than is possible with mixed-phase EoSs, as we describe in Section~\ref{quarkdec}.

The lack of a density regime with zero speed of sound for EoSs with percolation (but instead with some kind of structure) increases the stability of neutron stars against gravitational collapse. Additionally, it has been shown by many different studies that structure in the speed of sound of dense matter is expected from theory (see, e.g. \cite{Gusakov:2014ota,Stone:2019blq,Motornenko:2019arp,Baym:2019iky,McLerran:2018hbz,Zhao:2020dvu,Duarte:2020xsp,Sen:2020qcd,Ferreira:2020kvu,Hippert:2021gfs,Pisarski:2021aoz,Marquez:2022gmu,Albino:2024ymc,Jakobus_2021,Dexheimer_2015,Alford_2017,Zacchi_2016,Alvarez-Castillo:2018pve,Blaschke_2020,Dutra_2016,Xia_2021,Yazdizadeh_2019,Shahrbaf_2020,Zhao:2020dvu,Lopes_2021,rho2021fractionalizedquasiparticlesdensebaryonic,Marczenko_2022,Kapusta_2021,Somasundaram_2022,Mukherjee_2017,Li_2020,Jin_2022,Lee_2022,Dexheimer:2009hi,Malfatti_2020,Tu_2022,Monta_a_2019,Jokela_2021,Jeong_2020,Sen_2021_2,Li_2018,kumar2023,Marczenko:2022jhl,Zuo_2022,braun2022zerotemperaturethermodynamicsdenseasymmetric,Ivanytskyi_2022,fraga2022,pinto2022,kumar2023_2,kouno2024,tajima2024,ye2025highdensitysymmetryenergy,Kov_cs_2022}) and necessary for describing modern nuclear and observational data \cite{Bedaque:2014sqa,Tan:2021ahl,Li_2020,wang2020exploringhybridequationstate,Dutra_2016,Minamikawa_2021,kojo2021qcdequationsstatespeed,Raduta_2021,Dutra_2012,Annala:2019puf,PhysRevC.103.045804,Tews:2018kmu,Fujimoto:2022ohj,Huang_2022,Han:2022rug,Liu_2023,yamamoto2023quarkphasesneutronstars,gao2024reconcilinghessj1731347constraints,Legred_2022,Altiparmak_2022,mroczek2023nontrivialfeaturesspeedsound,Cuceu_2025,xia2024astrophysicalconstraintsnucleareoss}.
The speed of sound is important in the context of binary neutron-star mergers, as it controls the stiffness of matter, i.e.~how the pressure responds to changes in density. This is particularly relevant in the merger and early post-merger phases, where the neutron star fluid is most dynamic. It is important, then, to qualify whether these changes in the structure of the sound speed are likely to be of import to current and future gravitational wave observatories, which can only be done through simulations. Changes in the EoS stiffness can affect not only the frequencies of gravitational waves observed, but also the type of remnant formed (hypermassive neutron star or black hole) and how delayed the collapse to a black hole might be, assuming that outcome (see, e.g.~\cite{Hotokezaka:2011dh,Bauswein:2012ya,Bauswein:2013jpa,Bauswein:2015yca,Bernuzzi:2015rla,Rezzolla:2016nxn}). 

Neutron-star merger simulations have already been performed with EoSs containing percolation. In particular, Ref.~\cite{Huang:2022mqp} performed merger simulations (of neutron stars with different masses and radii) using two EoSs, one stiffer and one softer. The novelty in our work comes from performing merger simulations of groups (three pairs and one quartet) of neutron star binaries that have approximately the same mass, radius, and tidal deformability. These groups are generated by EoSs that produce stellar families that pass through a recurring region in the mass-radius and mass-tidal diagrams. See \cite{Clevinger:2025acg} for a detailed explanation of how recurring regions can be found and how they can be used to produce different EoSs that reproduce a given observation of neutron star or neutron-star merger. In other words, our chosen star configurations in each group have approximately the same masses, radii and tidal deformability, but different EoSs. 

To produce these EoSs, we use different microscopic (hereafter original) EoSs, two versions of the Chiral Mean Field (CMF) model and the modified Density-Dependent (DD2) model combined with the Nambu–Jona-Lasinio (NJL) model. The three original descriptions already contain a first-order phase transition between hadronic and quark matter, and we use them in this work to construct EoSs with higher-order phase transitions. The original descriptions provide state-of-the-art relativistic EoSs with the relevant degrees of freedom (nucleons and light quarks), one of them also including hyperons, the lightest spin-3/2 baryons (the $\Delta$-baryons), and strange quarks. They are  one-dimensional tabulated EoSs from the CompStar Online Supernovae EoSs (CompOSE), which is an online EoS repository \cite{Oertel:2016bki,Typel:2013rza}. 
CompOSE EoSs are provided as tables that contain thermodynamic, compositional, microscopic and other data in a common format. The percolation is produced using our open-source code provided in \cite{Clevinger:2025acg}. Our BNS simulations are performed using the GR-Athena++ code for general-relativistic magnetohydrodynamics (GRMHD) \cite{Cook:2023bag,Daszuta:2024chz}. Temperature effects are added to the EoSs with percolation using a thermal gamma-law (first used in numerical relativity by \cite{Shibata:2005ss}, see also \cite{Janka:1993,Mosta:2013gwu,Lattimer:2015nhk,Raithel:2019gws,Figura:2020fkj}), and initial conditions are calculated using LORENE \cite{Gourgoulhon:2000nn}. Throughout this work we, unless otherwise stated, assume $c=G=k_B=1$. 

In Section \ref{eosform}, we summarize the original tabulated microscopic descriptions for neutron-star matter. Then, in Section \ref{quarkdec}, we lay out our percolation prescription. We present our complete EoSs and discuss how to generate recurring regions in mass-radius and mass-tidal deformability diagrams in Section \ref{eosres}. We present the merger simulations performed using groups of EoSs with percolation in Section \ref{mergerres}. Finally, we present our conclusions in Section \ref{conc}. 

\section{Microscopic Formalism}
\label{eosform}

In this work we make use of two microscopic models. The first one is described in detail in Ref.~\cite{Clevinger:2022xzl}. We focus here on two of the 16 EoSs (hybrid EoSs 2\footnote{CMF-2: \url{https://compose.obspm.fr/eos/268}} and 7\footnote{CMF-7: \url{https://compose.obspm.fr/eos/273}}) described in this reference. While EoS 2 contains nucleons (and electrons) and a first order phase transition to up and down quarks, EoS 7 also contains muons, hyperons, Delta baryons and a first order phase transition to matter that also contains strange quarks. Both of these are taken to be in the zero-temperature limit, and reproduce charge neutral, beta-equilibrated matter. 
These EoSs derive from the Chiral Mean Field (CMF) model, which is a relativistic mean field model that reproduces chiral symmetry restoration and quark deconfinement at high energies \cite{Dexheimer:2009hi}. The model is fitted to reproduce nuclear-physics data at saturation density, astrophysics observations, lattice, and perturbative QCD.  Here we use a new version of the CMF model that contains additional vector and isovector interactions and a modified description of quarks from \cite{Dexheimer:2020rlp}. This allows for stable hybrid stars that do not need to include a mixed phase, in addition to being smaller but still massive enough to reproduce $2\,M_\odot$ observations \cite{Fonseca:2021wxt,Antoniadis:2013pzd}. Additionally, as described in Ref.~\cite{Clevinger:2022xzl}, the bulk-matter EoSs are attached to crusts containing nuclei in a way that ensures the symmetry energy slope, $L$, is continuous across the crust-core boundary.

The second microscopic model comes from a Relativistic Density Functional (RDF 1.7) description from \cite{Bastian:2020unt}\footnote{RDF: \url{https://compose.obspm.fr/eos/170}}. It uses a two-phase approach, where the hadronic and quark EoSs are connected necessarily by a first-order phase transition. For the hadronic phase, the EoS comes from the DD2F model with nucleons (and electrons), which is the Density-Dependent model with experimental nucleon masses (DD2) with a flow constraint correction \cite{Typel:2005ba,Typel:2009sy,Alvarez-Castillo:2016oln}. Within this phase, there is a transition to uniform hadronic matter from nuclei that is implemented using excluded volume effects to suppress nuclei at higher densities \cite{Alvarez-Castillo:2016oln}. The Nambu-Jona-Lasinio (NJL) model with up and down quarks is used with a string flip model (SFM) term and vector interactions to describe the quark phase \cite{Kaltenborn:2017hus}.

\section{Quark Deconfinement}
\label{quarkdec}

To create a percolation around the first-order phase transition in the original EoSs described above, we use the method described in \cite{Kojo:2014rca}. The difference is that, instead of using a fixed expansion for the percolation, imposing a fixed number of boundary conditions, and making one choice for the EoS, in this work we expand on all these fronts.
We start by choosing two values of $n_\mathrm{B}$ to control the size of the percolation region: the density where the pure hadronic phase ends and the density at which the pure quark phase begins. Then, we find the corresponding baryon chemical potentials $\mu_{\mathrm{B},\mathrm{H}}$ and $\mu_{\mathrm{B},\mathrm{Q}}$ respectively. To calculate the pressure in the percolation region between the two phases, creating a new phase, we use a polynomial $\mathcal{P}$ of order $M$, given by
\begin{align}
    \mathcal{P}(\mu_\mathrm{B}) &= \sum_{m = 0}^{M} b_m (\mu_\mathrm{B})^m \nonumber \\ &= b_0 + b_1 \mu_\mathrm{B} + b_2 \mu_\mathrm{B}^2 + b_3 \mu_\mathrm{B}^3 + b_4 \mu_\mathrm{B}^4 + ...~.
    \label{percolation}
\end{align}
$\mathcal{P}(\mu_\mathrm{B})$ contains $M+1$ coefficients, which we constrain by imposing $M+1$ boundary conditions. 

The number of constraints we match from derivatives of original EoS at each boundary determines the order of phase transition we construct. Here, we focus on cases where the order of the phase transition is the same at both boundaries. The boundary conditions are as follows
\begin{align}
     P(\mu_\mathrm{B} = \mu_{\mathrm{B},\mathrm{H}}) &= \mathcal{P}(\mu_\mathrm{B} = \mu_{\mathrm{B},\mathrm{H}}), \nonumber \\ 
     \quad \left. \frac{\partial P}{\partial \mu_\mathrm{B}} \right|_{\mu_\mathrm{B} = \mu_{\mathrm{B},\mathrm{H}}} &= \left. \frac{\partial \mathcal{P}}{\partial \mu_\mathrm{B}} \right|_{\mu_\mathrm{B} = \mu_{\mathrm{B},\mathrm{H}}}, \nonumber \\ 
     \quad \left. \frac{\partial^2 P}{\partial \mu_\mathrm{B}^2} \right|_{\mu_\mathrm{B} = \mu_{\mathrm{B},\mathrm{H}}} &= \left. \frac{\partial^2 \mathcal{P}}{\partial \mu_\mathrm{B}^2} \right|_{\mu_\mathrm{B} = \mu_{\mathrm{B},\mathrm{H}}}, \nonumber \\ ... &= ...,
    \label{boundaries}
\end{align}
and similar for $\mu_\mathrm{B} = \mu_{\mathrm{B},\mathrm{Q}}$, 
where $\mu_{\mathrm{B},\mathrm{H}}$ and $\mu_{\mathrm{B},\mathrm{Q}}$ are the baryon chemical potentials at the hadronic or quark phase boundary of the percolation region and $P(\mu_\mathrm{B})$ is the corresponding pressure (all from the original microscopic models).
The n\textsuperscript{th} derivative of the pressure is the n\textsuperscript{th}-order susceptibility 
$\chi_n^\mathrm{B}=\partial^n P / \partial \mu_\mathrm{B}^n = \partial^{n-1} n_\mathrm{B} / \partial \mu_\mathrm{B}^{n-1}$.
We can vary the order of our phase transition by matching a different number of derivatives (or susceptibilities) at the boundaries. E.g., 0\textsuperscript{th}-order derivative conditions correspond to a first-order phase transition, 1\textsuperscript{st}-order derivative conditions gives rise to a second-order phase transition, 2\textsuperscript{nd}-order derivative conditions gives rise to a third-order phase transition (with each case fitting 2, 4 and 6 total conditions respectively), etc. Additional conditions necessarily require higher-order terms in the polynomial to keep the number of constraints and coefficients, $b_m$ the same. This ensures a unique solution for the coefficients of $\mathcal{P}$. 
Alternatively, one can also impose specific constraints (beyond those found in Eq.~\ref{boundaries}) at the boundaries, such as specifying any of the higher-order derivatives, $\chi_{n}^\mathrm{B}$. In this work we do exactly that, and specify second-order derivatives $\chi^\mathrm{B}_2=\partial^2 P / \partial\mu_\mathrm{B}^2$ in some cases, at both sides of the percolation. By doing that, we obtain third-order phase transitions. In other cases, we do not fix $\chi^\mathrm{B}_2$, but match them instead on both sides of the percolation and obtain second-order phase transitions. In both cases, our polynomials are of 5th order (with 6 coefficients).

Finally, we must also consider whether our newly constructed EoSs are physical. There are several thermodynamic and astrophysical constraints to consider, in addition to causality. Other constraints at either low or high density are already fulfilled by the original EoSs. We impose a limit on the speed of sound ($0< c_s^2 < 1$, i.e.~the sound speed must be real and causal), that the pressure must be continuous, and that $n_\mathrm{B}$ must be a convex function of $\mu_\mathrm{B}$, i.e.~$\chi^\mathrm{B}_2\ge 0$ (see Ref.~\cite{Aloy:2018jov} for a discussion of concave EoS in the context of gravitational waves). The main astrophysical constraint we impose is that the EoS must be able to produce a neutron star with mass ${\geq}2\,M_{\odot}$, to be consistent with astrophysical observation of heavy neutron stars \cite{Fonseca:2021wxt,Antoniadis:2013pzd}. This can be checked by solving the Tolman-Oppenheimer-Volkoff (TOV) equations \cite{Tolman:1939jz,Oppenheimer:1939ne} with our constructed EoS as an input. Other relevant astrophysical observables, such as radius or tidal deformability, can also put constraints on producing physical EoSs. However, since the uncertainties on those measurements are much larger, instead of enforcing a specific limit for them as we did for the mass, we only use them as a guide for the discussion of the stars we reproduce. For a complete discussion on dense matter and observations, see \cite{MUSES:2023hyz}.

\section{Results for the Equation of State}
\label{eosres}

\begin{figure*}[p!]
\centering
\includegraphics[width=0.95\linewidth]{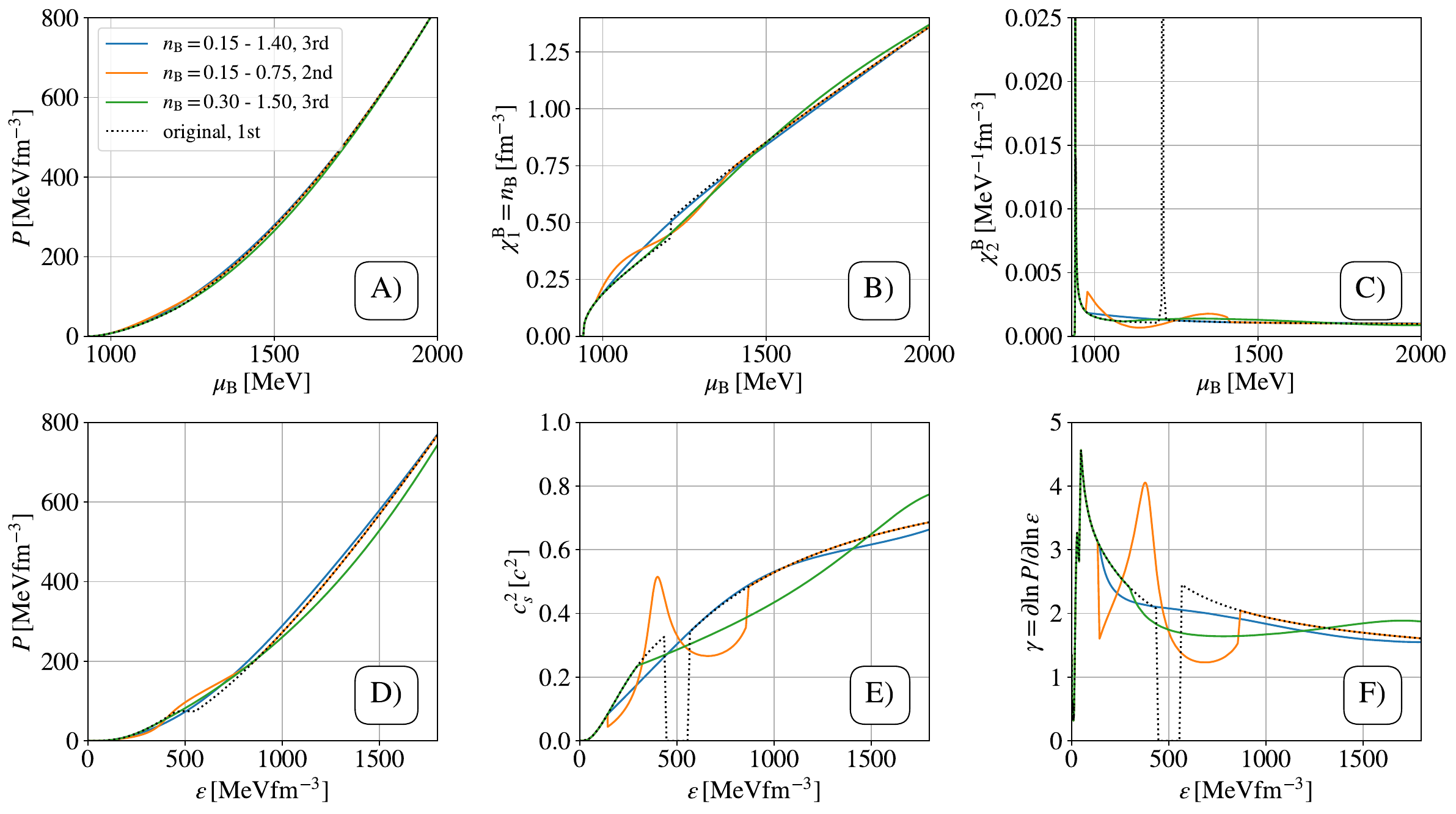}
\caption{Original microscopic CMF-2 EoS (black dotted line) and 3 different percolation parameter sets (colorful solid lines). The upper panels show pressure, $P$, (panel A), baryon number density, $n_\mathrm{B}$, (panel B), and the second derivative of the pressure with respect to the baryon chemical potential, $\chi^\mathrm{B}_2$, (panel C), all as functions of the baryon chemical potential, $\mu_\mathrm{B}$. The bottom panels show $P$ (panel D), the (squared) speed of sound, $c_s^2$ (panel E), and the adiabatic index, $\gamma$ (panel F), all as functions of the energy density, $\varepsilon$.}
\label{fig:EoS_cmf2}
\end{figure*}

\begin{figure*}[p!]
\centering
\includegraphics[width=0.95\linewidth]{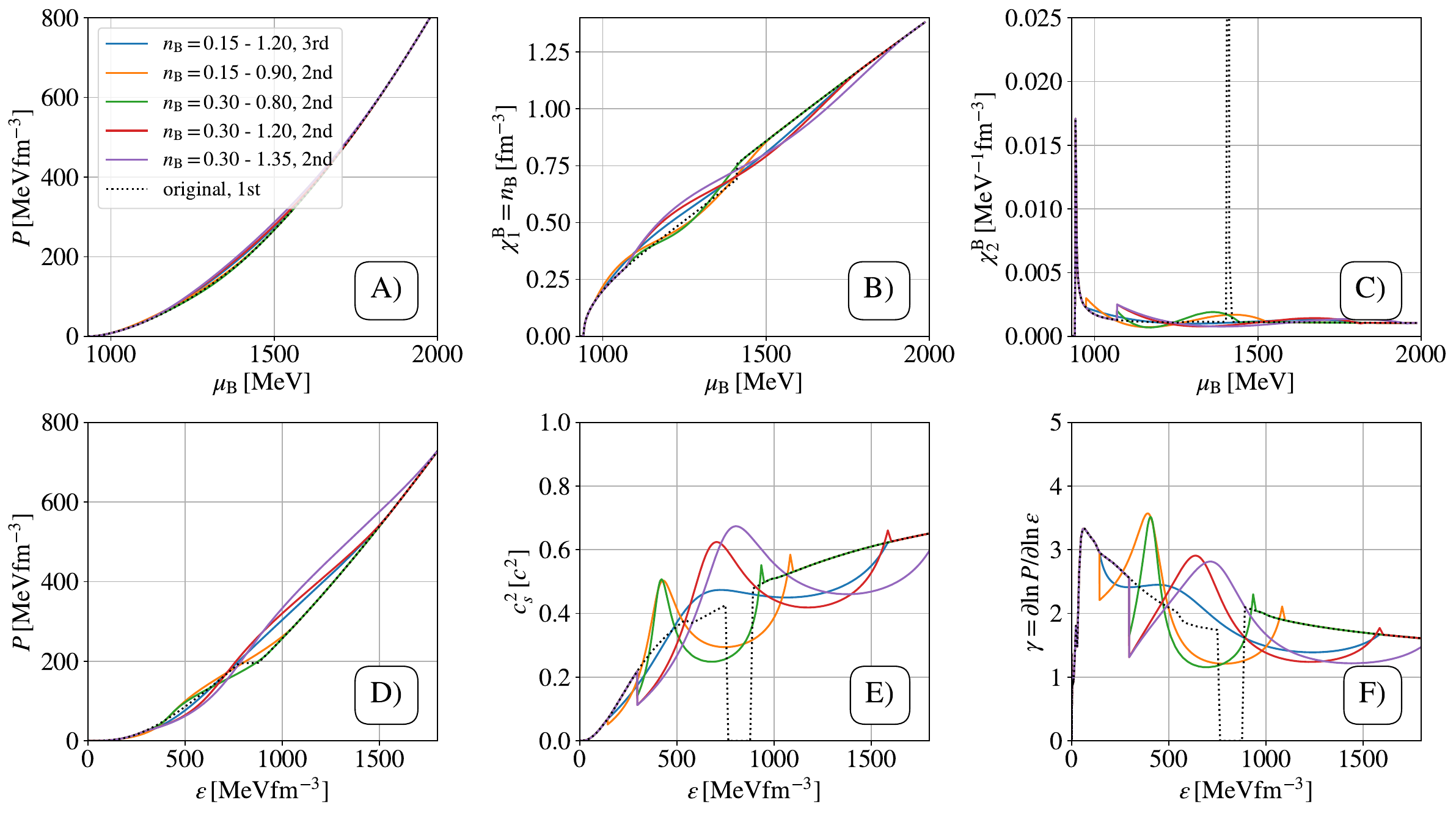}
\caption{Same as for Figure~\ref{fig:EoS_cmf2} but for microscopic EoS CMF-7 and 5 different percolation parameter sets.}
\label{fig:EoS_cmf7}
\end{figure*}

\begin{figure*}[p!]
\centering
\includegraphics[width=0.95\linewidth]{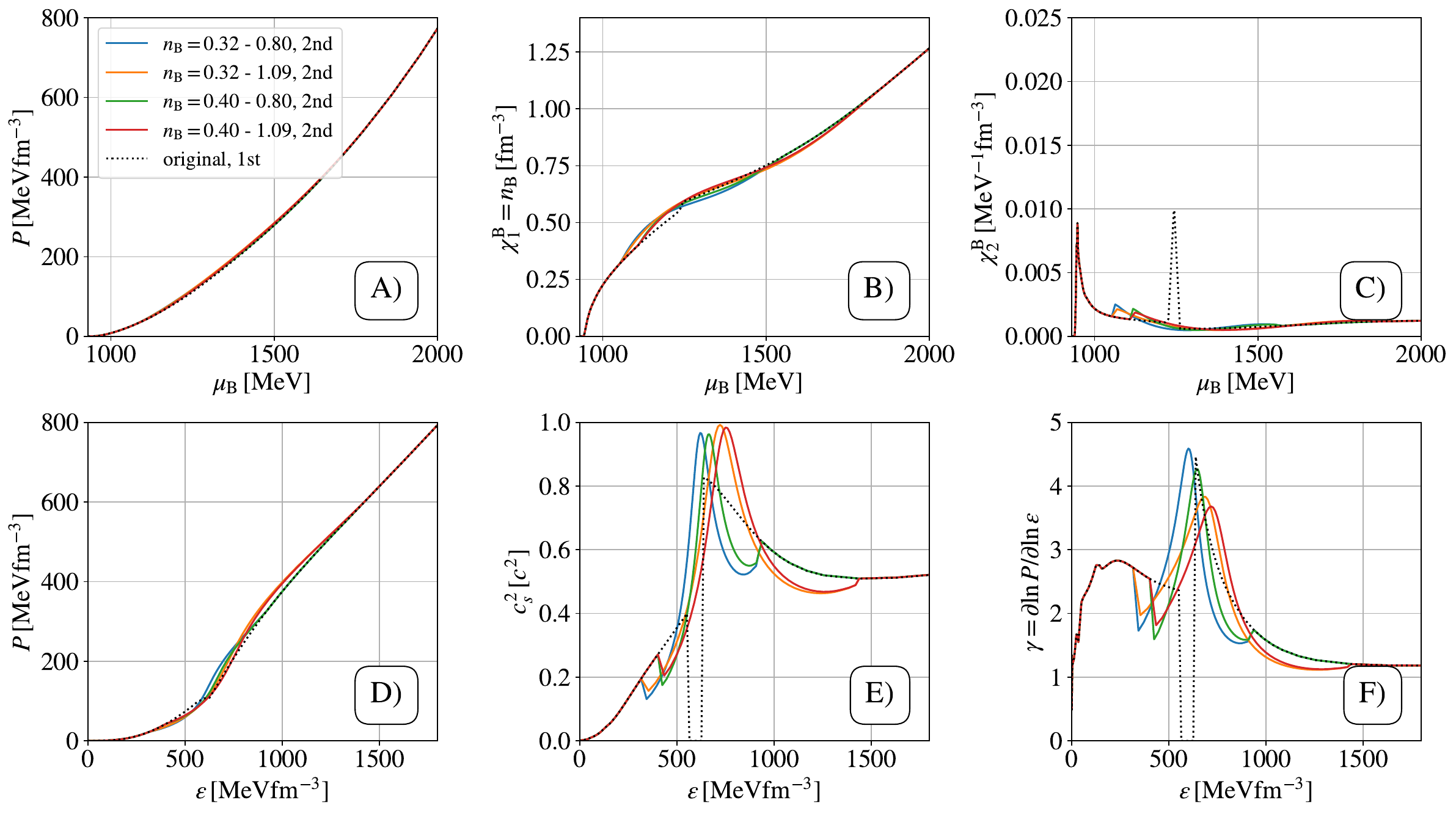}
\caption{Same as for Figure~\ref{fig:EoS_cmf2} but for microscopic EoS RDF 1.7 and 4 different percolation parameter sets.}
\label{fig:EoS_dd2f}
\end{figure*}

We present here our results for different original EoSs (shown as black dotted lines) and several percolation parameter sets for each EoS (shown as colorful solid lines).
In Figures~\ref{fig:EoS_cmf2} (nucleons and u,d quarks) and \ref{fig:EoS_cmf7} (nucleons, hyperons, $\Delta$'s and u,d,s quarks) we discuss the CMF EoSs (2 and 7 respectively) and the EoSs with percolation derived therefrom. The different panels show the pressure and its derivatives as a functions of different variables, including the speed of sound $c_s^2=\partial P/ \partial \varepsilon$ and adiabatic index $\gamma=\partial \ln P/ \partial \ln \varepsilon$. The differences are relatively small for the pressure as a function of $\mu_\mathrm{B}$ (panel A) but larger for pressure as a function of $\varepsilon$ and the derivatives thereof. A first-order deconfinement phase transition can be seen as a vertical and/or horizontal line for the original EoSs in all derivatives (and a zero sound speed). At very low densities, another line for the original EoSs can be seen in all derivatives, corresponding to the nuclear liquid-gas phase transition.

The percolation can smooth out the phase transition in different ways. In the case that the highest derivatives matched are the first ones, $\chi^\mathrm{B}_1$, this becomes a second-order phase transition (as indicated in the legends of Figures.~\ref{fig:EoS_cmf2} and \ref{fig:EoS_cmf7}, e.g.~the orange EoS in Figure~\ref{fig:EoS_cmf2}). In this case, there is a jump in $\chi^\mathrm{B}_2$ at both boundaries of the percolation region. We also study the case in which the highest derivatives matched are the second ones, $\chi^\mathrm{B}_2$, this becomes a third-order phase transition (e.g.~the blue EoS in Figure~\ref{fig:EoS_cmf2}). In this case there is no jump in $\chi^\mathrm{B}_2$. In addition to different orders, we also explore different initial and final densities for the percolation, $n_{\mathrm{B},\mathrm{H}}$ and $n_{\mathrm{B},\mathrm{Q}}$.
In Figure~\ref{fig:EoS_dd2f} we discuss the RDF 1.7 EoS and the EoSs with percolation derived therefrom. The results are equivalent to the CMF results, with the exception that, because it is harder for RDF to create physical percolations that match the constraints we impose, and observational data (they tend to become acausal), we only report on matches up to second order. This generates, as previously discussed, discontinuities in the second derivative $\chi^\mathrm{B}_2$.

\begin{figure*}[p!]
\centering
\includegraphics[width=0.95\linewidth]{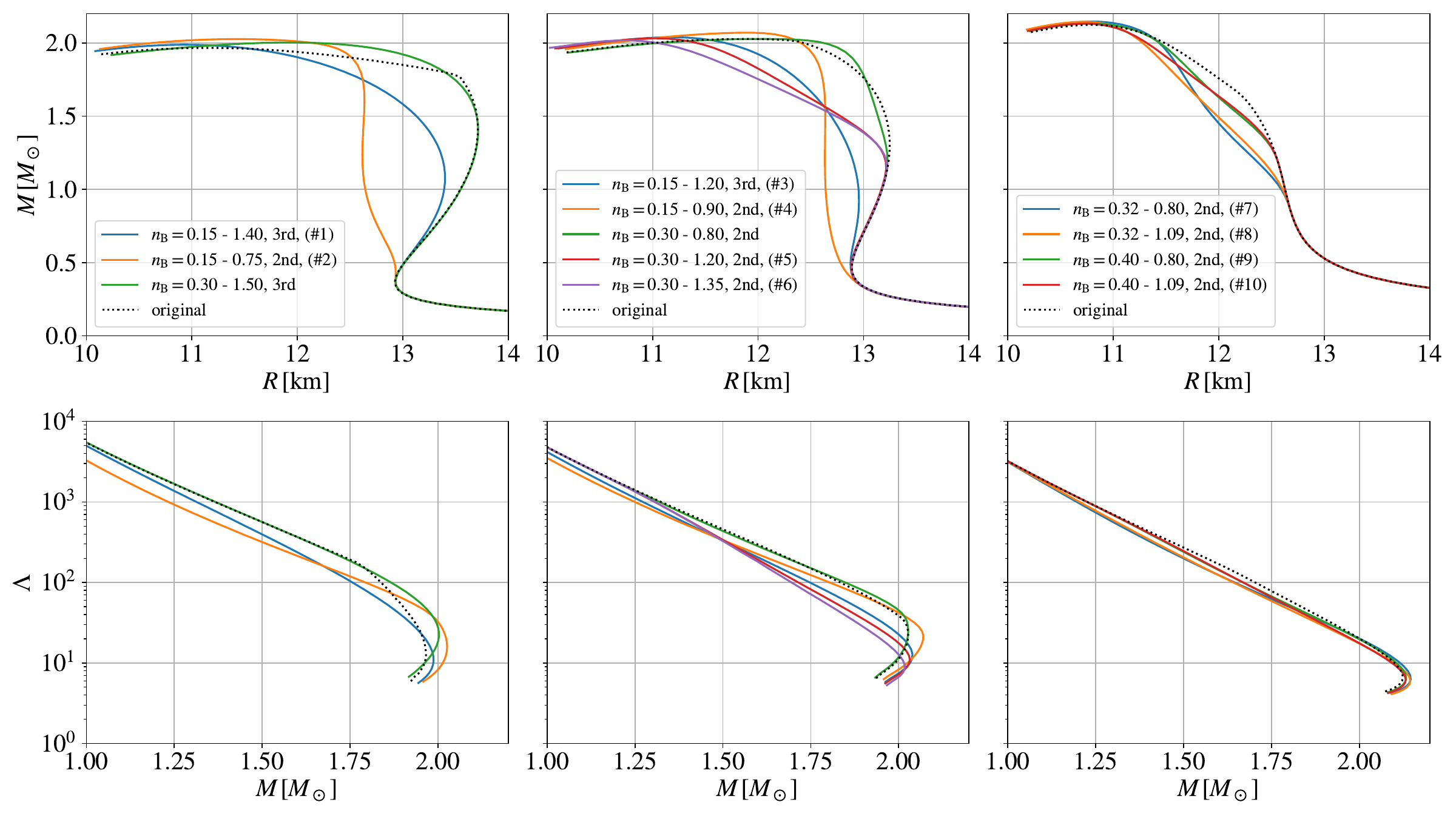}
\caption{Stellar properties for the different EoSs shown in Figure~\ref{fig:EoS_cmf2} for CMF-2, in Figure~\ref{fig:EoS_cmf7} for CMF-7, and in Figure~\ref{fig:EoS_dd2f} for RDF 1.7. The upper panels show gravitational mass, $M$, versus radius, $R$, and the lower panels show the dimensionless quadrupolar tidal deformability, $\Lambda$, versus $M$.}
\label{fig:TOV_dd2f}
\end{figure*}

Due to the observations of ${\ge}2\,M_\odot$ neutron stars \cite{Antoniadis:2013pzd,Fonseca:2021wxt}, we expect the dense matter EoS to be stiff. On the other hand, recent measurements of neutron-star radius \cite{Miller:2019cac,Riley:2019yda,Capano:2019eae,Miller:2021qha,Riley:2021pdl,Raaijmakers:2021uju,Vinciguerra:2023qxq,Salmi:2024aum,Salmi:2024bss,Choudhury:2024xbk,Mauviard:2025dmd} and, in particular tidal deformability \cite{LIGOScientific:2018cki,Radice:2018ozg,Coughlin:2018fis,De:2018uhw,Malik:2018zcf,Dietrich:2020efo,Chatziioannou:2020pqz,Breschi:2021tbm}, pointed to a softer EoS, or a smaller speed of sound at low densities. Additionally, perturbative QCD predicts a small speed of sound at asymptotically large densities \cite{Fraga:2013qra}. When combined, these constraints point to structure (more precisely a bump) in the speed of sound \cite{Tan:2020ics,Tan:2021ahl,Blaschke_2020,McLerran:2018hbz,Xia_2021,Duarte:2020xsp,Hippert:2021gfs,Pisarski:2021aoz,Sen:2020qcd,Kapusta_2021,Lee_2022,Bedaque:2014sqa,Tu_2022,Baym:2019iky,Tews:2018kmu,Fujimoto:2022ohj,braun2022zerotemperaturethermodynamicsdenseasymmetric,Liu_2023,Kawaguchi_2024,ye2025highdensitysymmetryenergy,Kov_cs_2022}. In a model-agnostic study, Bayesian inference was used to analyze the effects of imposing different constraints \cite{Gorda:2022jvk}. By combining the observational data with perturbative QCD, they also obtained a bump in the speed of sound. This speed-of-sound structure can be found in other model-independent studies \cite{Annala:2023cwx,Marczenko:2022jhl,Fujimoto:2022ohj,mroczek2023nontrivialfeaturesspeedsound,Altiparmak_2022,Han:2022rug}. This is exactly what we find, that the bump generated by the percolation can generate more massive and smaller stars (than the original EoSs), as shown in the mass-radius curves in the top panels of Figure~\ref{fig:TOV_dd2f}. 

We note that the curves for several of these percolations intersect at specific small regions in Figure~\ref{fig:TOV_dd2f}; in the top panels for mass-radius, as well as in the bottom panels, which show mass versus the dimensionless quadrupolar tidal deformability, $\Lambda$ (see, e.g., Eq.~3 of \cite{Chatziioannou:2020pqz}, hereafter `tidal deformability' for brevity). This is especially visible for the CMF-7 EoSs (in the middle panels), where this occurs near a mass of $1.5\,M_\odot$ and radius of $12.8\,\mathrm{km}$. We refer to this point as a recurring region, which we discuss in detail in a separate publication \cite{Clevinger:2025acg}. To summarize said work, given a chosen stellar mass, no matter the size and position of the percolation region, as well as the order of the phase transitions around it, the radius for a recurring region can be calculated by minimizing the average sound speed (with respect to the susceptibilities at the boundaries) from the beginning of the percolation region to the central density of stars. This establishes a new connection between EoSs and stellar properties. 

The EoSs that do not minimize the speed of sound, do not pass by the recurring region, see e.g. the green line in the middle panels of Figure~\ref{fig:TOV_dd2f}.
As can be seen in the left and right panels of Figure~\ref{fig:TOV_dd2f}, percolations constructed using the other microscopic EoSs also present recurring regions. For CMF-2, we show a recurring region at a mass of ${\sim}1.7\,M_\odot$ and radius of ${\sim}12.7\,\mathrm{km}$, whereas for RDF 1.7, we show two pairwise crossings at ${\sim}1.6\,M_\odot$ and ${\sim}11.8\,\mathrm{km}$, and ${\sim}1.7\,M_\odot$ and ${\sim}11.9\,\mathrm{km}$.
Furthermore, EoSs that pass by (or give rise to) a recurring region are very interesting for neutron-star merger simulations. As we discuss in the following, they produce nearly indistinguishable inspiral gravitational wave signals, allowing for direct comparison of the post merger signals. We focus on the 10 EoSs listed in Table~\ref{tab:sim_eoss} (all the ones that pass though recurring regions) from here on.

\section{Results from Merger Simulations}
\label{mergerres}

A series of binary neutron-star merger simulations were conducted using the state-of-the-art GR-Athena++ \cite{Cook:2023bag,Daszuta:2024chz} GRMHD code. The spacetime was evolved according to the Z4c formulation \cite{Bernuzzi:2009ex,Ruiz:2010qj,Weyhausen:2011cg,Hilditch:2012fp} of the Einstein field equations, the fluid is evolved according to the Valencia formulation \cite{Marti:1991wi,Banyuls:1997zz}, and an oct-tree AMR method \cite{Stone:2020,Daszuta:2021ecf} was used to obtain sufficient resolution close to the stars. Our simulations did not include magnetic fields. Initial data were produced using LORENE \cite{Gourgoulhon:2000nn} with gravitational masses listed in Table~\ref{tab:sim_eoss} and mass ratios of 1. For the evolution, we used a resolution of $184\,\mathrm{m}$ on the finest grid, with a CFL number (grid size to time step ratio) of $0.25$. A total of ten different simulations were completed, two utilizing CMF-2 EoSs with percolation (both merging stars from the recurring region), four for CMF-7 EoSs with percolation (all four merging stars from the recurring region), and four for RDF EoSs with percolation 1.7 (two merging stars from one recurring region and two from another recurring region). The details of these choices are included in Table~\ref{tab:sim_eoss}. All stars at the recurring regions have central densities within the respective percolation region of the EoS that generates them.

We focus on recurring regions as they can represent neutron star binaries with different EoSs but the same values for stellar mass, radius, and tidal deformability. The tidal deformabilities of the two stars determine the leading order correction to the gravitational wave phase calculated under the assumption of point masses \cite{Flanagan:2007ix,Hinderer:2007mb}, so matching both the mass and tidal deformability between pairs of binaries results in inspiral waveforms that are difficult to distinguish \cite{Hinderer:2009ca,Damour:2012yf,Read:2013zra}. Therefore, by choosing to focus on such regions, we can restrict the expected differences between the waveforms to the merger and postmerger only, allowing us to better see how the dynamics of the fluid are affected by the differences in our EoSs.
Note that recurring regions for the EoSs with percolation can be produced for any chosen stellar masses, although the corresponding radii must be consistently determined (the opposite could of course be constructed, but masses and radii of recurring regions cannot both be chosen, unless the original EoS is modified) \cite{Clevinger:2025acg}. 

\begin{table*}[p!]
\hspace{0.2cm}
\begin{tabular}{cc|ccccc|cccc|cc}
 Micro & Simulation & PT & $n_{\mathrm{B},\mathrm{H}}$ & $n_{\mathrm{B},\mathrm{Q}}$ & $\chi^{\mathrm{B}}_{2,\mathrm{H}}$ & $\chi^{\mathrm{B}}_{2,\mathrm{Q}}$ & $M$ & $R$ & $\Lambda$ & $n_{\mathrm{B},\mathrm{c}}$ & $M_\mathrm{max}$ & $n_{\mathrm{B},\mathrm{c},\mathrm{max}}$ \\
 EoS & Number & Order & $[\mathrm{fm}^{-3}]$ & $[\mathrm{fm}^{-3}]$ & $[\mathrm{fm}^{-3} \mathrm{MeV}^{-1}]$ & $[\mathrm{fm}^{-3} \mathrm{MeV}^{-1}]$ & $[M_\odot]$ & $[\mathrm{km}]$ & & $[\mathrm{fm}^{-3}]$ & $[M_\odot]$ & $[\mathrm{fm}^{-3}]$ \\ \hline
CMF-2 & 1 & 3 & 0.15 & 1.40 & $-$ & $-$ & 1.675 & 12.7 & 159.8 & 0.542 & 2.00 & 1.105 \\
CMF-2 & 2 & 2 & 0.15 & 0.75 & 0.0035 & 0.0015 & 1.675 & 12.7 & 159.9 & 0.453 & 2.03 & 0.956 \\ \hline
CMF-7 & 3 & 3 & 0.15 & 1.20 & $-$ & $-$ & 1.500 & 12.8 & 333.0 & 0.470 & 2.04 & 1.026 \\
CMF-7 & 4 & 2 & 0.15 & 0.90 & 0.003 & 0.001 & 1.500 & 12.8 & 338.5 & 0.421 & 2.07 & 0.877 \\
CMF-7 & 5 & 2 & 0.30 & 1.20 & 0.0025 & 0.001 & 1.500 & 12.8 & 338.6 & 0.508 & 2.03 & 1.062 \\
CMF-7 & 6 & 2 & 0.30 & 1.35 & 0.0025 & 0.001 & 1.500 & 12.8 & 328.1 & 0.528 & 2.02 & 1.124 \\ \hline
RDF 1.7 & 7 & 2 & 0.32 & 0.80 & 0.0028 & 0.0009 & 1.600 & 11.8 & 124.4 & 0.579 & 2.15 & 1.047 \\
RDF 1.7 & 8 & 2 & 0.32 & 1.09 & 0.0024 & 0.0012 & 1.600 & 11.8 & 123.8 & 0.607 & 2.14 & 1.121 \\
RDF 1.7 & 9 & 2 & 0.40 & 0.80 & 0.0024 & 0.0009 & 1.675 & 11.9 & 98.0 & 0.606 & 2.13 & 1.049 \\
RDF 1.7 & 10 & 2 & 0.40 & 1.09 & 0.0021 & 0.0012 & 1.675 & 11.9 & 97.7 & 0.626 & 2.13 & 1.122
\end{tabular}
\caption{Percolation parameter sets that reproduce groups of EoSs that go though recurring regions and are used to perform binary neutron star simulations. All binaries have a mass ratio of 1. The columns are: original microscopic EoSs, number of simulation, order of phase transition at percolation boundaries, densities at boundaries, susceptibilities at boundaries, stellar mass, radius, (dimensionless quadrupolar) tidal deformability, central density of stars at the recurring region, and maximum mass supported by the EoS and corresponding central density The dashed lines for $\chi^\mathrm{B}_2$ denote that these values were matched to the original EoS, not imposed as in other cases. While all stars at the recurring region have central densities inside of the percolation region, only some maximum-mass stars have large enough central densities to reach the quark phase.}
\label{tab:sim_eoss}
\end{table*}

In order to enable conservation of both baryon number and energy, the Valencia formulation \cite{Marti:1991wi,Banyuls:1997zz} used by GR-Athena++ requires at least two degrees of freedom in the EoS (without which the evolution of the conserved internal energy density, $\tau$, see Eq.~(22) of \cite{Marti:1991wi}, becomes redundant), so we add a temperature degree of freedom to the 1-dimensional barotropic EoSs described so far to approximate thermal effects.
This is achieved using a thermal gamma-law, based on the behaviour of an ideal gas \cite{Janka:1993,Shibata:2005ss}. The total pressure and energy density of the fluid are given by
\begin{align}
    \varepsilon \left(n_\mathrm{B}, T\right) &= \varepsilon_\mathrm{cold} \left(n_\mathrm{B}\right) + \varepsilon_\mathrm{th} \left(n_\mathrm{B}, T\right), \label{eqn:hybrid_energy} \\
    P \left(n_\mathrm{B}, T\right) &= P_\mathrm{cold} \left(n_\mathrm{B}\right) + P_\mathrm{th} \left(n_\mathrm{B}, T\right), \label{eqn:hybrid_pressure}
\end{align}
where the subscript cold terms refer to the EoSs as described above, and the thermal components (subscript th) are given by 
\begin{align}
    \varepsilon_\mathrm{th} \left(n_\mathrm{B}, T\right) &= \frac{n_\mathrm{B} T}{\Gamma_\mathrm{th} - 1}, \label{eqn:thermal_energy} \\
    P_\mathrm{th} \left(n_\mathrm{B}, T\right) &= n_\mathrm{B} T, \label{eqn:thermal_pressure} 
\end{align}
where the thermal adiabatic index $\Gamma_\mathrm{th}$ is a free parameter to be chosen, $T$ is the temperature, which, typically, is calculated from either the energy density or pressure and then used to determine the other. We can see that Eqs.~\ref{eqn:thermal_energy} and \ref{eqn:thermal_pressure} satisfy the ideal gas relation $p_\mathrm{th} = (\Gamma_\mathrm{th} - 1)\varepsilon_\mathrm{th}$, and accordingly we choose a value of $\Gamma_\mathrm{th} = 5/3$ to represent a monoatomic ideal gas (e.g.~non-relativistic free neutrons). The final ingredient is the sound speed with finite temperature (c.f.~definition in Section \ref{eosres}), which is given by 
\begin{align}
    c^2_s (n_\mathrm{B}, T) &= \frac{ \mathbbm{h}_\mathrm{cold}(n_\mathrm{B}) c^2_{s,\mathrm{cold}}(n_\mathrm{B}) + n_\mathrm{B} T}{ \mathbbm{h}_\mathrm{cold}(n_\mathrm{B}) +  \mathbbm{h}_\mathrm{th}(n_\mathrm{B}, T)}, \label{eqn:hybrid_cs2}
\end{align}
where $ \mathbbm{h}_\mathrm{\,x} = P_\mathrm{x} + \varepsilon_\mathrm{x}$ is the enthalpy density. 

For gravitational waves, our simulations provide the (forth) Weyl Scalar, $\Psi_4 = -\ddot{h}_{+} + i\ddot{h}_\times$, from the Newman-Penrose formalism of GR \cite{Newman:1961qr}, where $h_{+,\times}$ are the two gravitational wave polarisations (not to be confused with enthalpy density, $\mathbbm{h}$). This is decomposed into spherical harmonics, such that 
\begin{align}
    \Psi_4 (\theta,\phi) = \sum_{l, m} \Psi_{4}^{l, m}~{}^{-2}Y_{l, m} (\theta,\phi),
\end{align}
where $\Psi_{4}^{l, m}$ are the spin-weighted spherical harmonics of $\Psi_4$, and ${}^{s}Y_{l,m}$ are the corresponding weights \cite{Goldberg:1966uu}. We sum over modes from $l=2$ to $l=4$ and $m=-l$ to $m=l$, and we assume $\theta=\phi=0$ (i.e.~the detector is aligned with the spin axis of the binary). The time integration of $\Psi_4$ is performed using the Fixed-Frequency Integration technique of \cite{Reisswig:2010di}.

\begin{figure*}[p!]
\centering
\includegraphics[width=.75\linewidth,trim={0 1cm 0 0},clip]{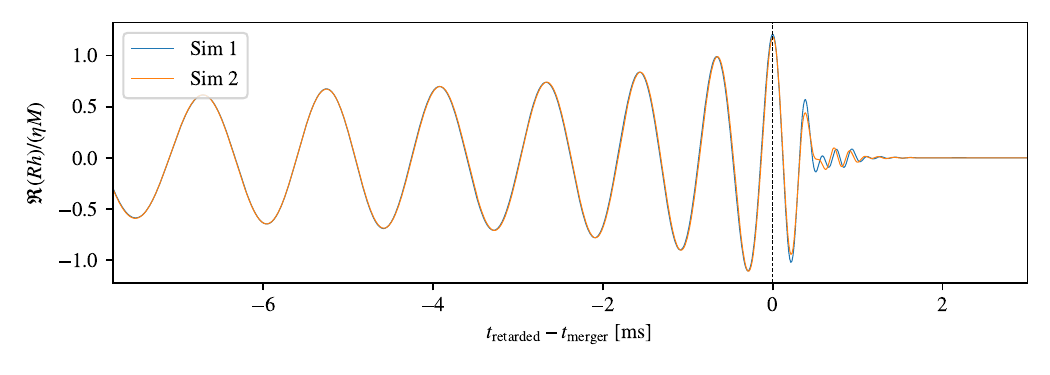}
\includegraphics[width=.75\linewidth,trim={0 1cm 0 0},clip]{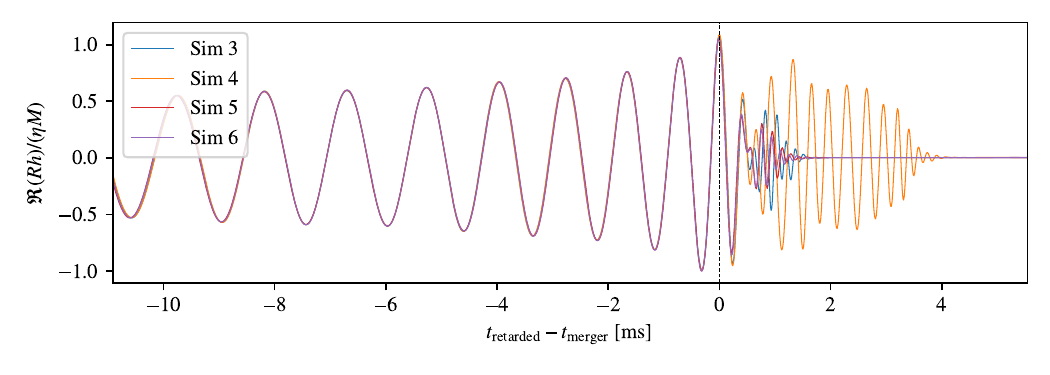}
\includegraphics[width=.75\linewidth]{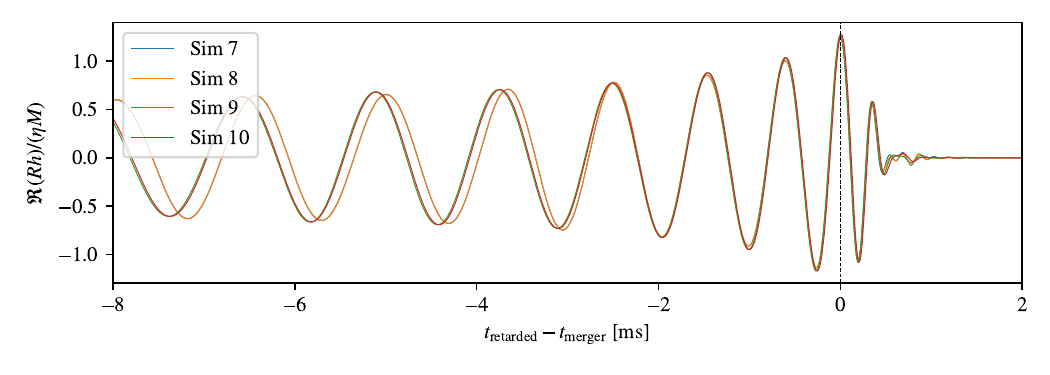}
\caption{Real part of the gravitational wave strain, $h$, for CMF-2 (upper panel), CMF-7 (middle panel), and RDF 1.7 (lower panel) following the numbering from Tab.~\ref{tab:sim_eoss}. Colours are chosen to match those used in Figures.~\ref{fig:EoS_cmf2}, \ref{fig:EoS_cmf7}, \ref{fig:EoS_dd2f}, and \ref{fig:TOV_dd2f}. Amplitudes are scaled by detector distance, $R$, the total mass of the system, $M=m_1 + m_2$ (calculated for each binary), and the symmetric mass ratio $\eta = m_1 m_2 / M^2$, and within each panel the signals are aligned in time and phase with the first listed simulation (blue in each panel) to minimise the pairwise mismatch calculated for inspiral frequencies (see text), except for the bottom panel (RDF 1.7) where Sim 8 is aligned to Sim 7, and Sim 10 is aligned to Sim 9.}
\label{fig:strain_comparison}
\end{figure*}

In Figure~\ref{fig:strain_comparison}, we see the real part of the gravitational wave strains, $h=h_+ - ih_\times$,
for the 10 simulations listed in Table~\ref{tab:sim_eoss}, grouped by original microscopic EoS. In the top panel, we see that both CMF-2 EoSs result in similar waveforms, as expected from their pairing in $M$ and $\Lambda$ (exact mismatches are discussed below). In the lower panel (RDF 1.7 EoS), we see that each pair of simulations, 7 \& 8 (blue and orange) and 9 \& 10 (green and red), give well matched waveforms, again as would be expected. However, in the middle panel (CMF-7 EoS), we see that three of the simulations, numbers 3, 5, and 6 (blue, red, and purple), collapse promptly after merger, as with the simulations for the other models whereas simulation 4 (orange) results in a short lived hypermassive neutron star that collapses on a time scale of ${\sim}3-4\,\mathrm{ms}$. Looking closer, we see more differences between those CMF-7 EoSs that do collapse promptly with this model, unlike in the other models where the post-merger behaviour is much more similar within an EoS group. The major contributing factor at play here is that the chosen recurring region takes place at a lower mass for the CMF-7 model (${\sim}1.5\,M_\odot$) than for CMF-2 (${\sim}1.7\,M_\odot$) and RDF 1.7 (${\sim}1.6-1.7\,M_\odot$). This lower mass allows the dynamics of the merger to be more affected by the differences in the EoSs, rather than a very immediate collapse to a black hole. For the remainder of this Section we will focus on the differences between the results of the CMF-7 simulations.

\begin{figure}[t!]
\centering
\includegraphics[width=.95\linewidth]{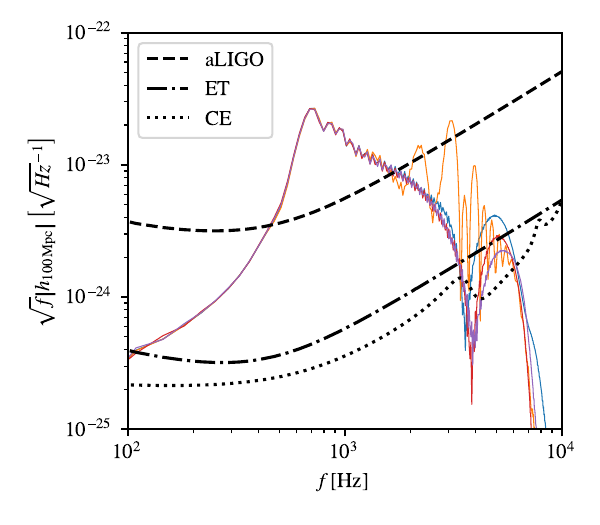}
\caption{Characteristic gravitational wave strain for simulations performed using the last group of EoSs from Tab.\ref{tab:sim_eoss} for CMF-7 model. Colours match those in Figure~\ref{fig:strain_comparison}. Amplitudes are scaled to a detector distance of $100\,\mathrm{Mpc}$, and we assume the event is directly overhead the detector, and at an inclination of $\theta=0$. Also included are the sensitivity curves for aLIGO, Einstein Telescope (ET-D), and Cosmic Explorer (CE2silica). Detector sensitivity curves for aLIGO and Cosmic Explorer were produced using pygwinc \cite{gwinc:2017}, and the Einstein Telescope curve was obtained from \cite{etd:2020}.}
\label{fig:strain_asd}
\end{figure}

In Figure~\ref{fig:strain_asd} we show the characteristic strain of the gravitational waves from the four simulations performed using the CMF-7 group of EoSs. As previously discussed, we see that simulations 3, 5, and 6 show the typical ringdown behaviour of a prompt collapse to a black hole, whereas simulation 4 (orange line) exhibits significantly more structure in the post-merger frequencies ($\gtrsim 2\,\mathrm{kHz}$) as a hypermassive neutron star is formed, completes several rotations, and then collapses to a black hole.

In order to quantify the differences between these signals, we can study the pairwise mismatches $\mathcal{M}$ \cite{Owen:1995tm} (see also, e.g.,~\cite{Lindblom:2008cm,Sathyaprakash:2009xs,McWilliams:2010eq,Baird:2012cu,DelPozzo:2014cla,Kumar:2015tha}). The mismatch is obtained from the faithfulness, $\mathcal{F}$, through
\begin{align}
    \mathcal{M} &= 1 - \mathcal{F}, \label{eqn:mismatch}
\end{align}
where $\mathcal{F}$ is itself defined as the maximised overlap between the two signals, accounting for both phase and time shifts ($\phi_c$ and $t_c$ respectively). The overlap $\left\langle \tilde{h}_1(f) | \tilde{h}_2(f)\right\rangle$ is defined as \cite{Finn:1992wt,Cutler:1994ys}
\begin{align}
    \left\langle \tilde{h}_1(f) | \tilde{h}_2(f) \right\rangle &= 4 \int_0^\infty \frac{\tilde{h}^\ast_1(f) \tilde{h}_2(f)}{S_n(f)} \mathrm{d} f, \label{eqn:overlap}
\end{align}
where $S_n(f)$ is the sensitivity of a given detector, and the tilde denotes a Fourier transform. From this we calculate $\mathcal{F}$ as \cite{Owen:1995tm}
\begin{align}
    \mathcal{F} &= \max_{\phi_c,t_c} \frac{\left\langle \tilde{h}_1(f) | \tilde{h}_2(f) e^{i(\phi_c - 2\pi f t_c)}\right\rangle}{\sqrt{\left\langle \tilde{h}_1(f) | \tilde{h}_1(f) \right\rangle \left\langle \tilde{h}_2 (f) | \tilde{h}_2 (f) \right\rangle}}. \label{eqn:match}
\end{align}

As, in principle, the waveforms for a real binary neutron-star merger would extend much longer before the merger itself than is currently possible to simulate numerically, we split the mismatch into two parts, pre- and post-merger, where only the post-merger part is of particular relevance here, although we can use the pre-merger mismatch to verify that the inspiral signals we do have match well in the way they should, as previously discussed. We can split the signal into pre- and post-merger frequencies by measuring the instantaneous frequency at $t_\mathrm{merger}$, $f_\mathrm{merger}$, where $t_\mathrm{merger}$ is the instant of maximum gravitational wave amplitude. Then, when calculating the faithfulness, we limit the integral in Eq.~\ref{eqn:overlap} to the frequency interval $0 \leq f \leq f_\mathrm{merger}$ for the pre-merger, and $f_\mathrm{merger} \leq f \leq \infty$ for the post-merger. We obtain $\phi_c$ and $t_c$ for each pair of signals by maximising the faithfulness over pre-merger frequencies, as the inspiral of any real signal would dominate the signal-to-noise ratio, though here it is truncated due to the finite length of the numerical relativity simulations.

\begin{figure}[t!]
\centering
\includegraphics[width=.95\linewidth]{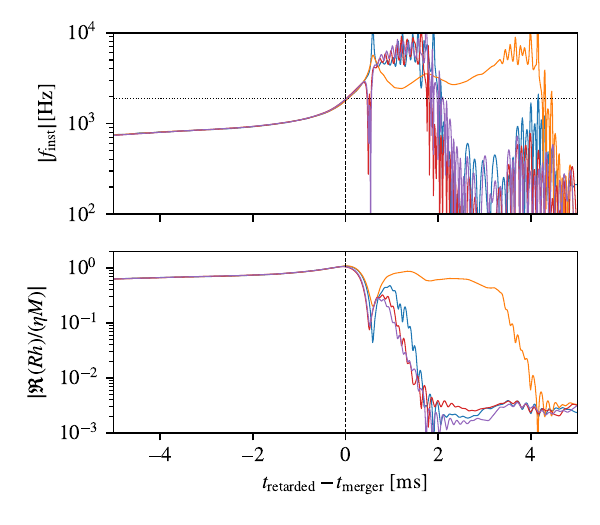}
\caption{Instantaneous frequency, $f_\mathrm{inst}$,(top) and gravitational wave strain amplitude, $h$, (bottom) for the CMF-7-group simulations. Colours match those in those in Figures.~\ref{fig:strain_comparison} and \ref{fig:strain_asd}. Amplitudes are scaled to a detector distance of $100\,\mathrm{Mpc}$. Vertical dashed line corresponds to the instant of maximum gravitational wave strain amplitude. Horizontal dotted line on upper panel is at $f_\mathrm{merger}=1.9\,\mathrm{kHz}$.}
\label{fig:finst}
\end{figure}

In Figure~\ref{fig:finst} we plot the absolute value of the instantaneous frequency, $\left|f_\mathrm{inst}\right| = (1/2\pi)\left|\partial \phi_\mathrm{GW} / \partial t\right|$, where $\phi_\mathrm{GW} = -\arg(h)$
is the phase of the gravitational wave strain, and the strain amplitude $\left| h \right|$. From this plot we see that the instantaneous frequency of all four signals is approximately $f_\mathrm{merger}=1.9\,\mathrm{kHz}$ at the point of maximum amplitude, hence we use this figure as the divide between pre- and post-merger gravitational waves.

\begin{figure}[t!]
\centering
\includegraphics[width=.9\linewidth]{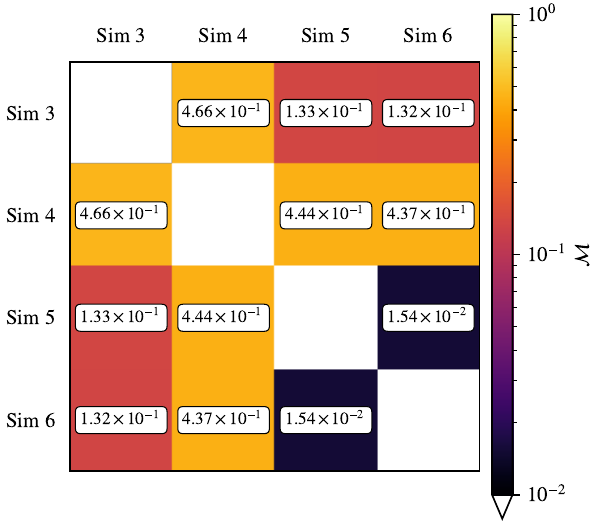}
\caption{Pair-wise post-merger mismatches for the CMF-7 group of simulations. A frequency of $f_\mathrm{merger}=1.9\,\mathrm{kHz}$ is used to exclude inspiral frequencies from the integral in Eq.~\ref{eqn:overlap} used in the mismatch calculation.}
\label{fig:mismatch}
\end{figure}

In Figure~\ref{fig:mismatch}, we give the measured pair-wise mismatches between all four EoSs with percolation from the CMF-7 group for the post-merger frequencies of the signal, using $f_\mathrm{merger}=1.9\,\mathrm{kHz}$ as discussed above. In these calculations we use the Einstein Telescope-D strain sensitivity curve \cite{Hild:2010id,etd:2020} for the noise spectrum in Eq.~\ref{eqn:overlap}. The calculated mismatches for the inspiral signals are all of order $\mathcal{O}(10^{-3})$, however, as can be seen in Figure~\ref{fig:mismatch}, the post-merger mismatches are significant larger. Of the promptly collapsing remnants, simulations 5 and 6 are most similar, with a post-merger mismatch of ${\sim}0.015$, whereas the mismatch between these two and simulation 3 is ${\sim}0.13$. The largest mismatches are between simulation 4 and the other three, which are around ${\sim}0.45$, and are result of the longer lived remnant in that case.

The mismatch between two signals is related to the signal to noise ratio (SNR), $\varrho$ (not to be confused with density, $\rho$) of their difference, $\varrho_\mathrm{diff}$ (see, e.g., \cite{McWilliams:2010eq} where this quantity is $\sqrt{\left\langle \delta h | \delta h \right\rangle}$), assuming the signals have similar SNR, $\varrho_\mathrm{signal}$. We can approximate $\varrho_\mathrm{diff}$ through
\begin{align}
    \varrho^2_\mathrm{diff} &\approx 2 \varrho^2_\mathrm{signal} \mathcal{M}. \label{eqn:snr_diff}
\end{align}
For two signals to be distinguishable, i.e.~dissimilar enough that an observed signal could be characterised as being of one kind or the other, we require that $\varrho_\mathrm{diff} \geq 1$. We can therefore rearrange Eq.~\ref{eqn:snr_diff} and impose this inequality to obtain
\begin{align}
     \varrho_\mathrm{signal} &\gtrsim \frac{1}{\sqrt{2 \mathcal{M}}}, \label{eqn:snr_min}
\end{align}
where $\varrho_\mathrm{signal}$ is now the minimum SNR required for the two signals to be distinguishable. For the most dissimilar signals, Sim 3 and Sim 4, a post-merger mismatch of $\mathcal{M} \approx 0.466$ results in a minimum SNR of $\varrho_\mathrm{signal} \gtrsim 1.04$, which, with the Einstein Telescope-D configuration, would be possible out to a luminosity distance of ${\sim}800\,\mathrm{Mpc}$ (c.f.~the expected BNS observation horizon of ET at $z=2 \sim 3$ \cite{ET:2019dnz}, which corresponds roughly to a distance of ${\sim}1\times10^4\,\mathrm{Mpc}$). Such a low $\varrho_\mathrm{signal}$ means that if the post-merger GW signal is observed at all, these signals are sufficiently different that it would be possible to tell one model from the other. On the other hand, Sims 5 and 6 would require a post-merger SNR of $\varrho_\mathrm{signal} \gtrsim 5.70$, which would be possible at a luminosity distance of ${\sim}100\,\mathrm{Mpc}$ with ET-D.

\begin{figure}[t!]
\centering
\includegraphics[width=0.95\linewidth]{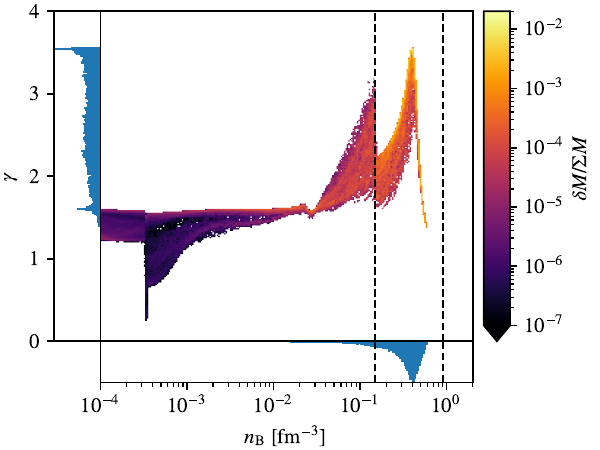}
\includegraphics[width=0.95\linewidth]{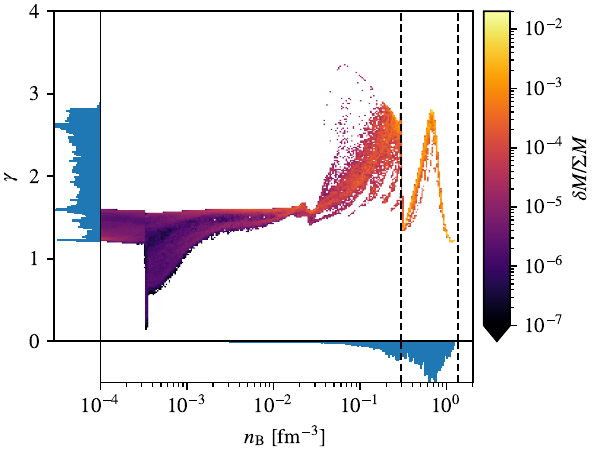}
\caption{2-Dimensional histograms of the density and temperature dependent adiabatic index for simulations 4 (upper panel) and 6 (lower panel), at the same time after merger (very shortly before the collapse of the remnant in simulation 6). The colormap denotes the amount of mass in each density-adiabatic index bin, $\delta M = \int \rho \sqrt{\det (\gamma^{ij})}  \, \mathrm{d}V$ (here $\gamma^{ij}$ is the spatial 3-metric of the fluid, not to be confused with the adiabatic index, and $\rho$ is the rest mass density), scaled to be relative to the total mass of the system $\Sigma M$. White indicates there is no matter corresponding to that bin. To the left and below each 2-D histogram is a 1-D histogram of the respective quantity. Dashed lines on the density axis denote the beginning and end of the percolation region for each EoSs. Lower densities are cut off as they do not contribute meaningfully to the total mass of the system.}
\label{fig:gamma_compar}
\end{figure}

To see why simulation 4 results in different behaviour to the other ones within the same group, it is informative to study how the stiffness of the EoS changes with density, and discuss how that affects the dynamics of the merger. In Figure~\ref{fig:gamma_compar} we plot histograms of the density versus the adiabatic index, which is calculated here using the temperature dependent equation (c.f.~definition in Section \ref{eosres})
\begin{align}
    \gamma (n_\mathrm{B}, T) &= \frac{\partial \ln(P(n_\mathrm{B}, T))}{\partial \ln(\varepsilon(n_\mathrm{B}, T))} = \frac{\varepsilon(n_\mathrm{B}, T)}{P(n_\mathrm{B}, T)} c_s^2 (n_\mathrm{B}, T),
\end{align} 
where $c_s^2 (n_\mathrm{B}, T)$ is calculated as in Eq.~\ref{eqn:hybrid_cs2}. The snapshots for simulations 4 and 6 (orange and purple respectively in Figures.~\ref{fig:EoS_cmf2}, \ref{fig:TOV_dd2f}, \ref{fig:strain_comparison},  \ref{fig:strain_asd}, and \ref{fig:finst}) are taken at the same time after merger, shortly before the remnant in simulation 6 collapses to a black hole. Both EoSs were built with percolation connected to the original hadronic and quark microscopic EoSs by second-order phase transitions. As we see by comparing the histograms, a significant amount of the matter in simulation 4 is at a higher adiabatic index than in simulation 6, while the density distribution peak occurs at lower density, $n_\mathrm{B,max} \approx 0.4\,\mathrm{fm}^{-3}$ versus ${\sim} 0.7\,\mathrm{fm}^{-3}$. This higher adiabatic index results in a larger pressure support against collapse in simulation 4, and correspondingly the remnant hypermassive neutron star is able to survive for several milliseconds, continuing to rotate and emit the gravitational waves we see in Figure~\ref{fig:strain_comparison}, before collapsing to a black hole. 

Comparing the values in Table~\ref{tab:sim_eoss}, we can see that Sim 4 has the lowest central density $n_\mathrm{B,c}$ in the CMF-7 group for both the recurring region star and the maximum mass star. This lower central density implies that more of the matter within the star---as compared to the other simulations---is stiffer, which correlates with what we see in Figure~\ref{fig:gamma_compar} (see Ref.~\cite{Tan:2021ahl} for a discussion on $c_s^2$ bumps and stellar properties).
The other simulations, particularly 5 and 6, also feature a bump in the sound speed and adiabatic index (as can be seen in Figure~\ref{fig:EoS_cmf7}), however, these bumps occur either at too high densities (simulations 5 and 6) or are not large enough (simulation 3) to meaningfully delay the collapse of the remnant. 

\section{Conclusions}
\label{conc}

In this work, we study the result of neutron-star mergers simulations using a collection of realistic neutron-star equations of state (EoSs), from crust to core, modeling deconfinement to quark matter as a higher-order (than first) phase transition. To produce many hybrid (hadronic and quark) EoSs in a consistent matter, we introduce a new intermediate phase of matter mimicking a quarkyonic phase \cite{McLerran:2007qj} (where quarks become deconfined but remain localized) by means of a percolation \cite{Kojo:2014rca}. Three different hybrid EoSs are obtained from CompOSE, all featuring a first-order deconfinement phase transition.  The pressure in the percolation region is calculated from a fifth-order polynomial, with six coefficients determined by matching the percolation boundaries to the original EoS phases using different orders of the pressure derivatives. More specifically, we either match derivatives of the pressure up to second-order (third-order phase transition) or match derivatives of the pressure only up to first-order (second-order phase transitions, imposing the second-order derivatives of the pressure - or susceptibilities). Although obtaining even higher orders for the deconfinement phase transition and mixed orders (different orders on each side of the percolation region) are possible, we found that these options were much more likely to produce acausal or non-monotonic EoSs, or EoSs that could not produce neutron stars with the required observed masses.

We focused on comparing EOSs that are causal, convex, and reproduce massive ($>2\,M_\odot$) stars, in addition to passing by recurring regions in the mass-radius and mass-tidal deformability diagrams. See \cite{Clevinger:2025acg} for details on how to produce such regions. We chose 10 configuration to run simulations: one group with 2 simulations using different percolation parameter sets for the CMF-2 model (both performed merging stars from the recurring region), one group with 4 simulations using different percolation parameter sets for the CMF7- model (all merging stars from the recurring region), and one group with 4 simulations using different percolation parameter sets for the CMF-2 model (merging stars from 2 recurring regions), as described in Tab.~\ref{tab:sim_eoss}. 

Each simulation was performed using the GR-Athena++ code for general-relativistic magnetohydrodynamics (GRMHD) \cite{Cook:2023bag,Daszuta:2024chz} with a thermal component added to the cold EoSs based on an ideal gas \cite{Janka:1993,Shibata:2005ss}. We used equal-mass binaries with initial data chosen such that (within the same group) the neutron stars had the same mass, radius, and tidal deformability, but different EoSs in the percolated region. This, in principle, should result in (almost) identical inspiral gravitational wave signals for the matching simulations, which is what we observe in the strain we extract from our simulations. 

In the post-merger gravitational wave signals, we see that most of the chosen simulations collapse promptly to form a black hole, as the masses of the recurring regions are relatively large ($1.5 - 1.675\,M_\odot$). However, one of the simulations, Sim 4, using a CMF-7-based EoS, resulted in a hypermassive neutron star that survived for ${\sim}3\,\mathrm{ms}$ before collapsing to a black hole. This difference resulted in a significant mismatch $\mathcal{M} \approx 0.45$ in the post-merger gravitational wave signal between this and the other simulations using the same original microscopic model, which had the same mass, radius, and tidal deformability for the initial binary configuration. We estimate that the pair of simulations with the greatest post-merger mismatch in the CMF-7 group (Sim 3 and Sim 4) would be distinguishable by the Einstein Telescope-D configuration out to $800\,\mathrm{Mpc}$, whereas the most similar pair in the same group (Sim 5 and Sim 6) would only be distinguishable out to $100\,\mathrm{Mpc}$ with the same observatory. 

Comparing the longer lived simulation (Sim 4) to one of the promptly collapsing simulations from the same group (Sim 6), we see that there is a significant amount more matter with higher adiabatic index $\gamma$ at the same time after merger (shortly before the collapse of the other simulation to a black hole). This stiffer matter is better able to resist the gravitational collapse of the star due to the increased pressure support. Note that, while most of the EoSs studied here do feature a bump in speed of sound, $c_s^2$, and $\gamma$, it generally occurs at too high density to meaningfully impact the dynamics of the merger.

This difference between post-merger behaviour using EoSs and binary setups that result in very similar inspirals raises an interesting challenge for the reconstruction of the full EoS from only inspiral data (as is possible with the current generation of gravitational wave observatories). As the dense matter portion of the EoS, especially concerning how quarks are deconfined, is still largely unknown, only an analysis of the post-merger part of gravitational wave signals from neutron star mergers will be able to break such degeneracy.

Additionally, there are many other effects currently under investigation by the community that result in differences in the post-merger gravitational wave signal, as will be targeted by next generation detectors, and disentangling these effects will pose a not-insignificant challenge once binary neutron star post-merger gravitational wave signals become available.
The simulations performed in this work represent only an exploratory study into the qualitative differences in observables that can be obtained by modifying the structure of quark deconfinement phase transition (while producing neutron stars with the same masses, radii, and tidal deformabilities). Future work could be done to study the effects of unequal mass binary systems and neutron star spins, as well as more consistent description of temperature effects (see e.g., \cite{Bauswein:2010dn,Raithel:2019gws,Perego:2019adq,Carbone:2019pkr,Figura:2020fkj,Hammond:2021vtv,Raithel:2021hye,Fields:2023bhs,Mroczek:2024sfp}) and out of $\beta$-equilibrium effects \cite{Alford:2018lhf,Alford:2019qtm,Endrizzi:2019trv,Most:2022yhe,Espino:2023dei,Yao:2023yda}. Additionally, one could construct a family of similar EoSs with a systematically varying bump in $c_s^2$ to study quantitatively how this affects the transition from prompt collapse to short-lived hypermassive neutron star.

\section*{Acknowledgments}

This work is partially supported by the NP3M Focused Research Hub supported by the National Science Foundation (NSF) under grant No. PHY-2116686.
VD also acknowledges support from the Fulbright U.S. Scholar Program and the Department of Energy under grant DE-SC0024700.
JF and DR acknowledge funding from U.S. Department of Energy, Office of Science, Division of Nuclear Physics under Award Number DE-SC0021177.
DR acknowledges funding from the National Science Foundation under Grants No.~PHY-2020275, AST-2108467, PHY-2407681, the Sloan Foundation, and from the U.S. Department of Energy, Office of Science, Division of Nuclear Physics under Award Number and DE-SC0024388.
MA expresses sincere gratitude to the FCT for their generous support through Ph.D. grant number \href{https://doi.org/10.54499/2022.11685.BD}{2022.11685.BD}. MA  and CP acknowledge the partial supported by funds from FCT (Fundação para a Ciência e a Tecnologia, I.P, Portugal) under projects UIDB/04564/2020 and UIDP/04564/2020, with DOI identifiers 10.54499/UIDB/04564/2020 and 10.54499/UIDP/04564/2020, respectively.
SB and BD acknowledges support by the EU Horizon under ERC
Consolidator Grant, no. InspiReM-101043372. 
The numerical simulations were performed on TACC's Frontera under NSF LRAC allocation PHY23001.

\bibliography{apssamp}

\end{document}